# High-Dimensional Light Field Modulation via Conjugate Phase Encoding in Liquid Crystal Devices

Tian Xia[#], Quanzhou Long[#], Wanlong Zhang*, Zhenwei Xie* and Xiaocong Yuan*

*Nanophotonics Research Center, Institute of Microscale Optoelectronics & State Key Laboratory of Radio Frequency Heterogeneous Integration, Shenzhen University, Shenzhen 518060, China*

[#]These authors contributed equally to this work.

*E-mail: zwl@szu.edu.cn; ayst31415926@szu.edu.cn; xcyuan@szu.edu.cn

**Abstract:**

High-dimensional light field modulation demands precise control over multiple optical parameters, a capability critical for next-generation photonic systems. While liquid crystals offer inherent advantages in dynamic birefringence tuning, existing approaches face fundamental limitations in decoupling interdependent phase responses across polarization states. Here, we demonstrate a conjugate phase encoding paradigm enabling simultaneous manipulation of wavelength-dependent wavefronts, orbital angular momentum (OAM), and polarization states in photoaligned liquid crystal devices. Through conjugate phase engineering, our single-element liquid crystal device generates 216 distinct holographic channels - a 6×6×6 matrix spanning spectral ($\Delta\lambda$=3 nm), OAM, and polarization ($\Delta\theta$=2°) dimensions. Using the proposed multiplexed lens—capable of simultaneous multi-spectral, multi-polarization, multi-spiral, and multi-focus beam control—we demonstrated liquid-crystal modulation and achieved high-capacity, multidimensional information multiplexing across wavelength, polarization, and OAM. This advancement establishes a transformative framework for ultrahigh-density optical information systems, achieving an unprecedented combination of spatial mode density and polarization multiplexing efficiency. The demonstrated capabilities significantly expand the design paradigm for multifunctional photonic devices, with direct applicability in holographic data storage architectures, polarization-encoded optical communications, and multidimensional optical cryptography systems. Our findings provide both methodological innovation and practical implementation strategies for sophisticated light field manipulation across multiple physical dimensions.

**Introduction**

The growing demand for advanced photonic devices necessitates the manipulation of propagating light across multiple degrees of freedom, including amplitude, phase, polarization state, and frequency. This capability is essential for achieving desired functionalities in applications such as optical sensing, imaging, and communication [1–5]. Unlike traditional bulk refractive optics, planar optical elements enable light manipulation based on fundamental principles of diffraction, effectively dispersing and redirecting light beams [6–8]. Through numerical calculations and simulations, diffractive optical elements can be designed with intricate microstructures on their surfaces, allowing for compact and lightweight configurations while facilitating multifaceted optical field manipulation [9]. The ability to generate precisely structured light with multiple degrees of freedom opens up possibilities for multiplexed information processing, significantly enhancing information capacity in optical storage and encryption [10,11]. More recently, as a subclass of two-dimensional optical metamaterials, optical metasurfaces present a promising approach to manipulate optical fields at subwavelength scales, offering advantages such as compactness and ease of integration [12–17]. By adjusting the geometrical parameters of these metasurfaces, one can modulate the phase delay of diffracted light, demonstrating the capability to control wavefronts with precision [18,19]. This versatility allows metasurfaces to function similarly to traditional optical components in various applications, including imaging [20,21], sensing [22], holography [23], and nonlinear processes [24]. Despite these advancements, challenges such as high fabrication costs, optical efficiency, and large panel sizes remain significant hurdles for the widespread adoption of optical metasurfaces.

Liquid crystals (LCs) have emerged as a popular class of nanomaterials due to their broad birefringence across ultraviolet to microwave wavelengths and their ability to be easily reconfigured under external fields, such as electric and thermal stimuli [25]. LC-based spatial light modulators are widely utilized to control the propagation phase of incident light by electrically adjusting the effective refractive index on a pixel-by-pixel basis, enabling various photonic applications [26–28]. Recent developments in photoinduced alignment technology have enhanced the orientation of multi-domain LC molecules at the microscale, allowing for the modulation of the geometric phase (also known as Pancharatnam-Berry (PB) phase) with high conversion efficiency [29,30]. These planar optical devices, featuring spatially varying LC alignment profiles, enable diverse beam shaping functions by altering the polarization-dependent PB phase [31,32]. Applications include beam steering for expanded eye boxes in augmented reality displays using polarization gratings [33], multi-focusing for camera imaging and augmented reality via PB lenses [34], and quantum optical processing employing $q$-plate vortex generators [35]. Furthermore, the photoinduced LC retarders are also proven to be applicable for all-optical neural networks and differential computation [36,37]. However, the limitation of a single geometric phase restricts the modulation capabilities, hindering control over LCs and ultimately constraining advancements in communication and information storage capacities.

In this study, we validate the concept of multifaceted optical field modulation through an innovative conjugate phase encoding method, introducing the LC orbital angular momentum (OAM) multiplexed holography technique based on the multiplexed lens (ML), which achieves simultaneous multi-spectral, multi-polarization, multi-spiral, and multi-focus beam control. Leveraging the unlimited orthogonal eigenstates of OAM, this technique can theoretically provide a vast number of transmission channels [38]. This approach enables the generation of multi-polarization, multi-color and multi-OAM holography, allowing precise control over wavefront, wavelength, and polarization using the geometric phase of a single LC retarder. The MLs facilitate the creation of distinct holographic images across a wide spectrum of incident wavelengths and OAM modes. Different OAM configurations produce various color images characterized by specific polarization states. The LC OAM multiplexed holography technique demonstrates the capability to encode 6 polarization states, 6 wavelengths, and 6 topological charges into a total of 216 multiplexed holograms. This results in exceptional multiplexing accuracies of 3 nm in wavelength differences and 2° in polarization azimuth variations. This advancement holds significant promise for applications in three-dimensional full-color displays, large-bandwidth information transmission, optical secure encryption, and other cutting-edge technologies.

## Results

**Design principle of conjugate phase encoding method on a single liquid crystal retarder**

The proposed LC retarder is designed to independently manipulate the wavelength, wavefront, and polarization of incident light through the encoding of conjugate phases. As a demonstration for large scale information encoding on a single LC retarder, Figure 1 illustrates the operational principle of the LC-based broadband OAM multiplexed holography. As light with varying wavelengths and wavefronts passes through the engineered LC retarder, distinct diffractive holographic patterns are projected onto the screen. For incident light with a specific topological charge, the horizontal rows of holographic patterns correspond to the diffraction patterns of individual wavelengths, while the vertical columns represent holographic patterns with desired arbitrary polarization states on the Poincaré sphere. Given that the diameter of the doughnut pattern of the incident beam is proportional to the topological charge, the MLs focus the incoming light and transform the OAM beam into Gaussian patterns. This results in the projection of different OAM-based holographic patterns along the longitudinal direction.

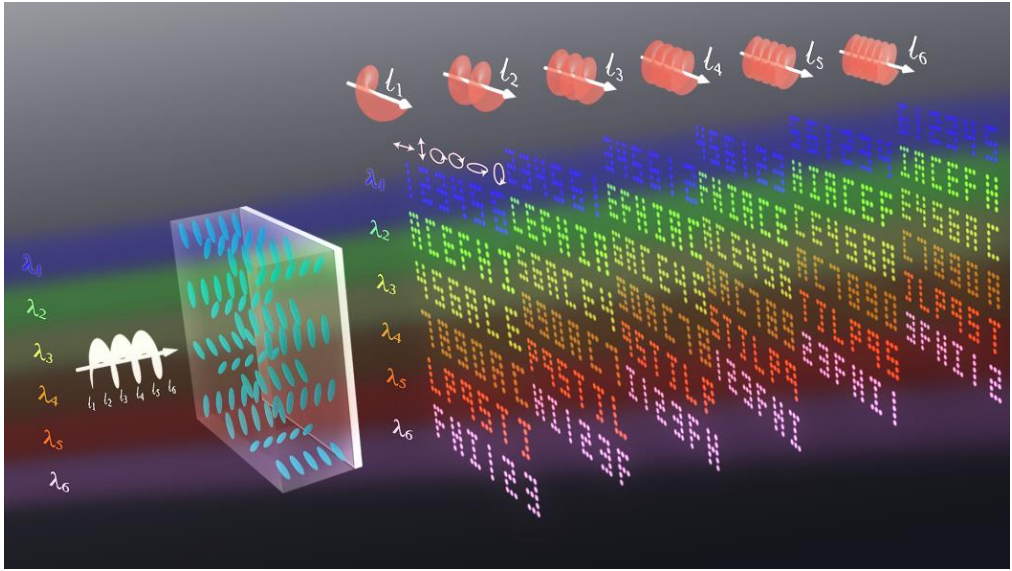

**Fig. 1** Operational principle of the LC-based broadband OAM multiplexed holography

The photoinduced alignment technology is applied to orient the LC molecules in plane, controlling the geometric phase of the incident light. The birefringence of LC molecules with different azimuthal angle can be described using the following Jones matrix:

$$J = \begin{bmatrix} \cos\frac{\Gamma}{2} - i\sin\frac{\Gamma}{2}\cos 2\theta & -i\sin\frac{\Gamma}{2}\sin 2\theta \\ -i\sin\frac{\Gamma}{2}\sin 2\theta & \cos\frac{\Gamma}{2} + i\sin\frac{\Gamma}{2}\cos 2\theta \end{bmatrix}, \quad (1)$$

where $\Gamma = 2\pi(n_e - n_o)d/\lambda$ represents the phase retardation, $\theta$ is the azimuthal angle of the LC director, $\lambda$ is the wavelength, $n_e$ and $n_o$ are the extraordinary and ordinary refractive indices respectively, and $d$ is the thickness of the LC layer.

With a linearly polarized light $E = A\exp(i\varphi)\begin{bmatrix}1\\0\end{bmatrix}$ propagates through the LC retarder, the transmissive optical field $f(\lambda,\varphi)$ can be represented as:

$$f(\lambda,\varphi) = \cos\frac{\Gamma}{2}A\exp(i\varphi)\begin{bmatrix}1\\0\end{bmatrix} - i\sin\frac{\Gamma}{2}\cos 2\theta A\exp(i\varphi)\begin{bmatrix}1\\0\end{bmatrix} + (-i\sin\frac{\Gamma}{2}\sin 2\theta)A\exp(i\varphi)\begin{bmatrix}0\\1\end{bmatrix}, \quad (2)$$

where $A$ and $\varphi$ denote the amplitude and phase of the incident beam, respectively.

Impose the geometric phase $2\theta$ as described in the following expression $\arg(\exp(i(\psi_1+\alpha))+\exp(i(\psi_2+\alpha)))$ and substitute it into Eq. (2). Moreover, a single phase can generate the complex amplitude based on double phase encoding. Thus, after setting $\exp(i\psi_1) = A_1\exp(i\phi(\lambda))$ and $\exp(i\psi_2) = A_2\exp(-i\phi(\lambda))$, where $A_1$ and $A_2$ are the amplitudes of the phases $\phi(\lambda)$ and $-\phi(\lambda)$ respectively, we can obtain the transmitted optical field $f(\lambda,\varphi,\alpha)$ as shown in:

$$f(\lambda,\varphi,\alpha) = f_1(\lambda,\varphi) + f_2(\lambda,\varphi,\alpha) + f_3(\lambda,\varphi,\alpha)$$
$$f_1(\lambda,\varphi) = \cos\frac{\Gamma}{2} A\exp(i\varphi)\begin{bmatrix}1\\0\end{bmatrix}$$
$$f_2(\lambda,\varphi,\alpha) = -i\frac{1}{\sqrt{2}}\sin\frac{\Gamma}{2} A\exp(i\varphi)[A_1\exp(i(\phi(\lambda)+\alpha))e_l + A_2\exp(i(\phi(\lambda)-\alpha))e_r]$$
$$f_3(\lambda,\varphi,\alpha) = -i\frac{1}{\sqrt{2}}\sin\frac{\Gamma}{2} A\exp(i\varphi)[A_2\exp(i(-\phi(\lambda)+\alpha))e_l + A_1\exp(i(-\phi(\lambda)-\alpha))e_r]$$
(3)

where $\alpha$ is the polarization azimuth for each holographic pattern, $e_l$ and $e_r$ are the left and right circular polarization basis vectors, respectively. Since the two phases $\phi(\lambda)$ and $-\phi(\lambda)$ corresponding to two elliptical polarization states are characterized with different azimuth angles ($\psi_1$, $\psi_2$) and ellipticity angles ($\chi_1$, $\chi_2$). Specifically, for phase $\phi(\lambda)$, $\psi_1$ and $\chi_1$ can be expressed as $\alpha$ and $\frac{1}{2}\arcsin\frac{A_1^2-A_2^2}{A_1^2+A_2^2}$, while for the reverse phase $-\phi(\lambda)$, $\psi_2$ and $\chi_2$ can be expressed as $\alpha$ and $\frac{1}{2}\arcsin\frac{A_2^2-A_1^2}{A_1^2+A_2^2}$. Moreover, it is important to note that given a lens phase profile $\phi(\lambda)$, the reverse phase $-\phi(\lambda)$ produces speckle background noise that does not affect the imaging quality of phase $\phi(\lambda)$ encoded light beam. The final polarization patterns can be regarded as generated solely by $\phi(\lambda)$, and the reverse phase encoded patterns can be disregarded. Similarly, for other phase profiles, although $-\phi(\lambda)$ cannot be entirely ignored as the background noise, the phases $-\phi(\lambda)$ and $\phi(\lambda)$ can be spatially separated to mitigate their mutual influence, thereby achieving independent control. As a result, the polarization pattern of phase $\phi(\lambda)$ remains unaffected by the optical field of reverse phase $-\phi(\lambda)$. It is noteworthy that the aforementioned amplitudes are not limited to constant values, and the double phase encoding can encode arbitrary complex amplitudes into the phases. This proposed conjugate phase encoding method thus enables arbitrary control over the wavelength, polarization, and wavefront of the geometric phase in a single LC retarder.

**Design principle of liquid crystal multifocal spiral lens for broadband orbital angular momenta multiplexing**

Owing to their unbounded helical modes and intrinsic orthogonality, OAM-multiplexed holography offers exceptional capacity for optical information processing[16]. The diameter of the doughnut pattern produced by a light beam is directly proportional to its OAM, meaning that beams carrying broadband OAMs will focus at varying distances when passed through a convex lens. This distinctive characteristic allows for the encoding of phase profiles with broadband OAMs, which can then be projected as holographic patterns at various planes using the lens. This capability not only enhances the versatility of holographic imaging but also significantly increases the information-carrying capacity of optical systems.

The phase profile of a typical lens that produces a distinct focal point can be described by the following equation:

$$\varphi(x, y) = -\frac{\pi}{\lambda f}[(x - x_0)^2 + (y - y_0)^2], \tag{4}$$

where $\lambda$ is the wavelength, $f$ is the focal length, $(x, y)$ are the coordinates on the incident plane, and $(x_0, y_0)$ are the coordinates of the focal point. When a beam passes through the lens with this phase profile, it generates a single bright spot on the focal plane, effectively reconstructing a holographic pattern. The foci can be arranged in an $n \times m$ array, and the phase profile of the multi-focus lens (MFL) can be calculated by:

$$\phi_{MFL}(x, y) = \arg\left\{\sum_{i=1}^{m}\sum_{j=1}^{n} e^{i\varphi_{i,j}(x,y)}\right\}. \tag{5}$$

For the MFL with a single focal plane, the phase of the lens at $(x_i, y_j)$ can be computed as $\varphi_{i,j}(x, y) = -\frac{\pi}{\lambda f}[(x - x_i)^2 + (y - y_j)^2]$. The coordinates of the focal point at $(x_i, y_j)$ on the focal plane are given by $x_i=(i-1-\text{floor}(m/2))d$ and $y_j=(\text{floor}(n/2)+1-j)d$, where $d$ is the distance between adjacent foci, corresponding to the sampling distance of the horizontal 2D Dirac comb, which is the $n \times m$ focal centers in the focal plane of the multifocal lens.

The phase profile for a typical spiral lens that creates a distinct focal point can be expressed as:

$$\psi(x, y) = -\frac{\pi}{\lambda f}[(x - x_0)^2 + (y - y_0)^2] + l\theta, \tag{6}$$

where $l$ is the topological charge and $\theta$ is the azimuthal angle. To further generate a patterned OAM holographic image, the $q$th row and the $p$th column 2D Dirac comb sampling point of the target image corresponds to the spiral foci of the spiral lens at $(x_p, y_q)$. The phase profile of the multi‑focus multi‑spiral lens that generates the sampling target image can be calculated by:

$$\phi_{MFL,l}(x, y) = \arg\left\{\sum\sum e^{i\psi_{p,q}(x,y)}\right\}. \tag{7}$$

When the multi‑spiral MFL has a single focal plane, the phase of the spiral lens at $(x_p, y_q)$ can be calculated as $\psi_{p,q}(x, y) = -\frac{\pi}{\lambda f}[(x-x_p)^2 + (y-y_q)^2] + l_1\theta$, where $l_1$ is the topological charge. Each pixel in the designed holographic image corresponds to the focal point at $(x_p, y_q)$ on the focal plane, defined as $x_p=(p\text{-}1\text{-floor}(m/2))d$ and $y_q=(\text{floor}(n/2)+1\text{-}q)d$.

To control the polarization state of each focal vortex for the multi-spiral MFL with incident linearly polarized light, the phase of the corresponding LC multi-polarization multi-spiral MFL can be calculated by:

$$\begin{aligned}\phi_{LCMFL,l,\psi,\chi}(x, y) &= \arg\left\{\sum\sum\{A_{1,p,q}e^{i[\psi_{p,q}(x,y)+\theta_{p,q}]} + A_{2,p,q}e^{i[-\psi_{p,q}(x,y)+\theta_{p,q}]}\}\right\} \\ &= \arg\left\{\sum\sum\{e^{i[\Psi_{1,p,q}(x,y)+\theta_{p,q}]} + e^{i[\Psi_{2,p,q}(x,y)+\theta_{p,q}]}\}\right\}\end{aligned}, \tag{8}$$

where $\theta_{p,q}$ is the azimuth angle of the elliptical polarization state of the focal vortex at $(x_p, y_q)$. $A_{1,p,q}$ and $A_{2,p,q}$ represent the encoded amplitudes of the phases $\psi_{p,q}(x, y)$ and $-\psi_{p,q}(x, y)$, respectively. Based on the double phase encoding, $\Psi_{1,p,q}(x, y)$ and $\Psi_{2,p,q}(x, y)$ are used to encode the complex amplitudes $A_{1,p,q}e^{i\psi_{p,q}(x,y)}$ and $A_{2,p,q}e^{-i\psi_{p,q}(x,y)}$, respectively.

The phase of the corresponding LC multi-spectral multi-polarization multi-spiral MFL under the wavelength $\lambda$ can be calculated by

$$\begin{aligned}\phi_{LCMFL,l,\psi,\chi,\lambda}(x, y) &= \arg\left\{\sum\sum\{A_{1,p,q,\lambda}e^{i[\psi_{p,q,\lambda}(x,y)+\theta_{p,q,\lambda}]} + A_{2,p,q,\lambda}e^{i[-\psi_{p,q,\lambda}(x,y)+\theta_{p,q,\lambda}]}\}\right\} \\ &= \arg\left\{\sum\sum\{e^{i[\Psi_{1,p,q,\lambda}(x,y)+\theta_{p,q,\lambda}]} + e^{i[\Psi_{2,p,q,\lambda}(x,y)+\theta_{p,q,\lambda}]}\}\right\}\end{aligned}, \tag{9}$$

where $\psi_{p,q,\lambda}(x, y)$ is the phase of the spiral lens at $(x_p, y_q)$ under the wavelength $\lambda$, and $\theta_{p,q,\lambda}$ is the polarization rotation angle of the focal vortex under the wavelength $\lambda$. $A_{1,p,q,\lambda}$ and $A_{2,p,q,\lambda}$ are the encoded amplitudes of the phases $\psi_{p,q,\lambda}(x, y)$ and $-\psi_{p,q,\lambda}(x, y)$, respectively. Based on the double phase encoding, $\Psi_{1,p,q,\lambda}(x, y)$ and $\Psi_{2,p,q,\lambda}(x, y)$ are used to encode the complex amplitudes $A_{1,p,q,\lambda}e^{i\psi_{p,q,\lambda}(x,y)}$ and $A_{2,p,q,\lambda}e^{-i\psi_{p,q,\lambda}(x,y)}$, respectively. Thus, the final phase information $\Phi$ of the LC ML under different wavelength channels can be obtained by:

$$\Phi = \arg\{\sum e^{i\phi_{LC\,MFL,l,\psi,\chi,\lambda}(x,y)}\}. \tag{10}$$

**Design principle of advanced holographic multiplexing with independent wavelengths, orbital angular momenta and polarization states**

The proposed conjugate phase encoding method facilitates the LC based ultra-high capacity holographic multiplexing with wavelengths, orbital angular momenta and polarization states, independently.

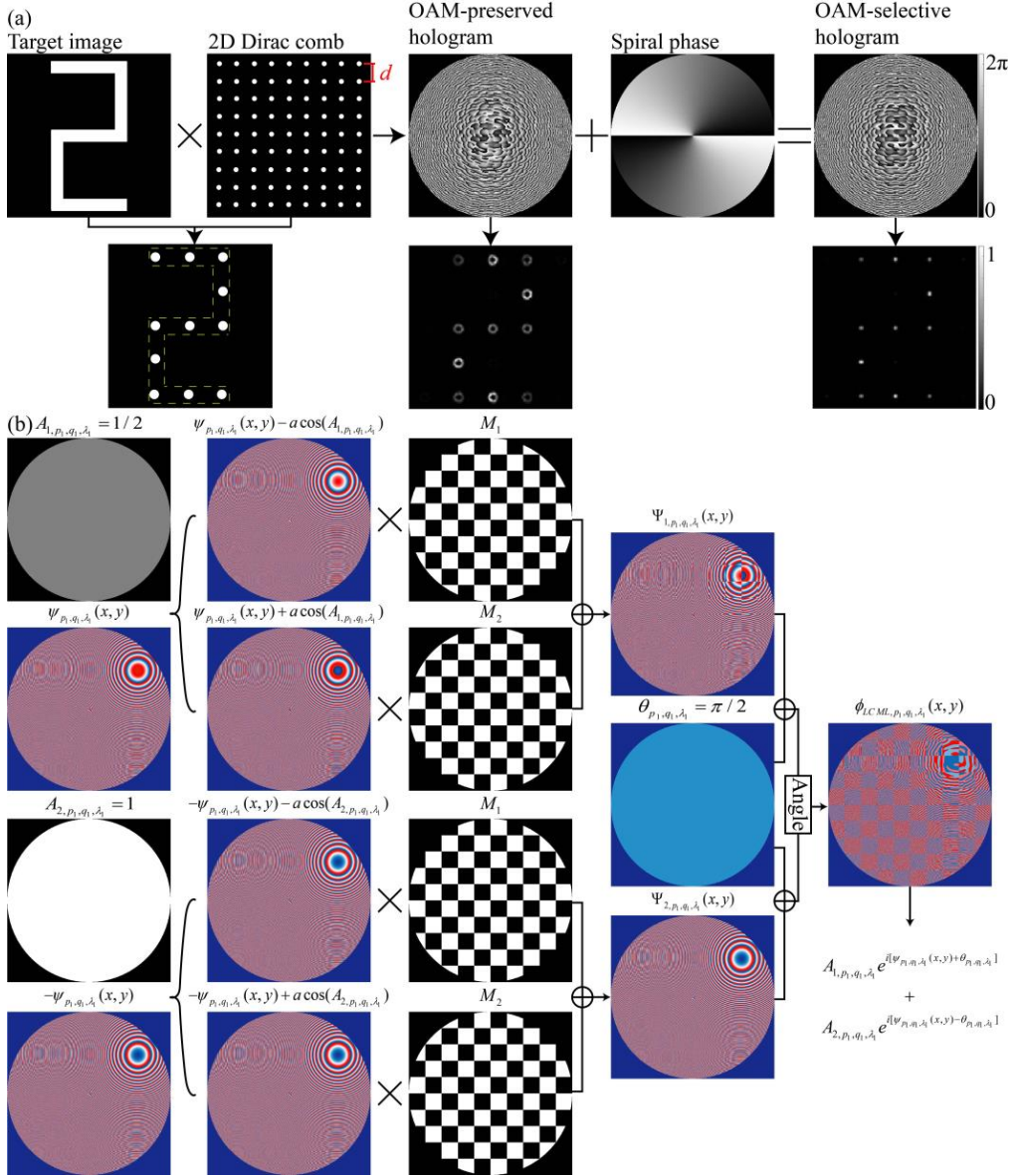

**Fig. 2** (a) Design concept for OAM-preserved and OAM-selective holograms. (b) the phase encoding procedure of the LC ML.

Figure 2(a) presents the design principle for the OAM-preserved and OAM-selective holography. The encoded target image is initially sampled with a 2D Dirac comb, with the distance between adjacent foci denoted as *d*, to create the designed spiral foci for the subsequent spiral lens. The middle section of Fig. 2(a) illustrates the phase profile of OAM-preserved hologram generated by employing the MLs. It is important to note that vortex beam is preserved in each pixel of the reconstructed image.

In the right portion of Fig. 2(a), we superimpose the phase function of an incident vortex beam with a topological charge (for instance, $l=-2$) onto the OAM-preserved hologram, resulting in the formation of an OAM-selective hologram. Due to the conservation of OAM, only a specific incident vortex beam with $l=-2$ can reconstruct Gaussian spots with a stronger intensity distribution in the desired holographic image.

Due to the high precision required for both wavelength and polarization angle, retrieving a decrypted image after lattice processing proves challenging. Consequently, additional image processing is performed on the lattice-filtered results. An intensity threshold $I_{Thr}$ is established such that if the intensity at a given point is greater than or equal to this threshold, the intensity of that point is set to 1, indicating the brightest intensity. Points that do not meet this criterion are filtered out and assigned an intensity of 0. Furthermore, since the effective area of the target image is defined as a 2D Dirac comb array, any high-intensity noise points generated outside this area due to optical field interference are also eliminated, resulting in an intensity of 0.

To illustrate the design process of LC based ultra-high capacity OAM-preserved holography, Fig.2(b) provides an example of polarization wavefront manipulation at a single point with a particular wavelength $\lambda_1$. The phase $\phi_{LCMSL,p_1,q_1,\lambda_1}(x,y)$ of the LC ML at $(x_{p_1}, y_{q_1})$ under the wavelength $\lambda_1$, which is based on the phase of the spiral lens located at the $q_1$th row and $p_1$th column, $\psi_{p_1,q_1,\lambda_1}(x,y)$, $-\psi_{p_1,q_1,\lambda_1}(x,y)$, along with the corresponding encoded amplitudes $A_{1,p_1,q_1,\lambda_1}=1/2$, $A_{2,p_1,q_1,\lambda_1}=1$, and the azimuthal angle $\theta_{p_1,q_1,\lambda_1}=\pi/2$ of the elliptical polarization state, is used to generate the focal vortex at $(x_{p_1}, y_{q_1})$ under the wavelength $\lambda_1$ with the elliptical polarization state of the azimuthal angle $\theta_{p_1,q_1,\lambda_1}$ and ellipticity angle $\frac{1}{2}\arcsin\frac{A_{1,p_1,q_1,\lambda_1}^2 - A_{2,p_1,q_1,\lambda_1}^2}{A_{1,p_1,q_1,\lambda_1}^2 + A_{2,p_1,q_1,\lambda_1}^2}$. Moreover, the complementary binary masks ($M_1$ and $M_2$), featuring 2D checkerboard patterns, are utilized to encode conjugate phases for achieving the manipulation of arbitrary complex amplitudes and polarization states. The phase generating the polarized focusing vortex at the other positions, along with the above phase, is utilized to construct the phase $\phi_{LCMSL,\lambda_1}(x,y)$ of the LC ML under the wavelength $\lambda_1$. The phase of the LC ML constructed at different wavelengths, follows the same methodology as described for the LC ML phase at the aforementioned wavelength $\lambda_1$. The final phase $\Phi$ displayed on the far right, is obtained by superimposing the phases of all the LC MLs across all wavelengths, as described in Eq. (10).

**Holographic multiplexing accuracies with individual wavelengths and polarization states**

The conjugate phased encoded LC retarder possesses the capability to manipulate optical field across multiple degrees of freedom, facilitating ultra-high-capacity multiplexed holography. It is crucial to explore the limitations of multiplexing accuracy for individual incident wavelengths and characterized polarization states.

When incident light with varying wavelengths interacts with the spiral lens, the corresponding focal points are distributed at slightly different positions, leading to corresponding holographic patterns that exhibit a slight shift along the propagation direction. Distinguishing between different holographic patterns on focal planes with closely spaced wavelengths can be challenging. In this study, the holographic patterns for different incident wavelengths are spatially arranged in distinct rows on the projection plane, as illustrated in Fig. 1.

To investigate the multiplexing limitations for incident wavelengths, the LC ML with a maximum radius of 5.832 mm and a focal length of 70 mm are designed to generate two different holographic images, "2" and "6," at wavelengths of 457 nm and 460 nm, respectively. The fabrication of the designed LC retarder for wavelength-only multiplexing followed the standard procedures outlined in the *Sample Fabrication* section. Fig. 3(a) displays the LC retarder optimized for the finest wavelength differences in holography, as observed through a polarization optical microscope, with a microscale multi-domain LC alignment depicted in Fig. 3(b). Fig. 3(c) presents both simulation and experimental results for the incident wavelengths of 457 nm and 460 nm. After applying the image processing procedures—lattice filtering, intensity threshold processing, and noise removal outside the effective area—it is evident that the holographic pattern "2" is projected on the left side with an incident wavelength of 457 nm, while the pattern "6" appears on the right side with an incident wavelength of 460 nm. The experiment further validates the presence of these two distinct numbers corresponding to their incident wavelengths. Notably, due to the extremely short focal length of the lens in the simulation, the holographic image consists of points with very few pixels, making it difficult to recognize. To address this, we apply a logarithmic transformation to these points and multiply them by a Gaussian function to generate larger Gaussian spots, thereby enhancing image recognition.

To minimize crosstalk caused by overlapping incident light spectra, the wavelength of 457 nm is generated using a single-frequency laser with a single longitudinal mode (MSL-R-457 from CNI, China), featuring a spectral linewidth of less than 0.00001 nm. The wavelength of 460 nm is produced using a narrow linewidth laser (MDL-E-460 from CNI, China) with a spectral linewidth of less than 0.03 nm. The right side of Fig. 3(c) shows the experimental holographic images captured by a CCD for each individual incident wavelength, where the corresponding patterns are spatially distributed approximately 1 mm apart. These clear results indicate that the wavelength multiplexing encoding method has been validated, achieving a remarkable precision of 3 nm in the wavelength spectrum.

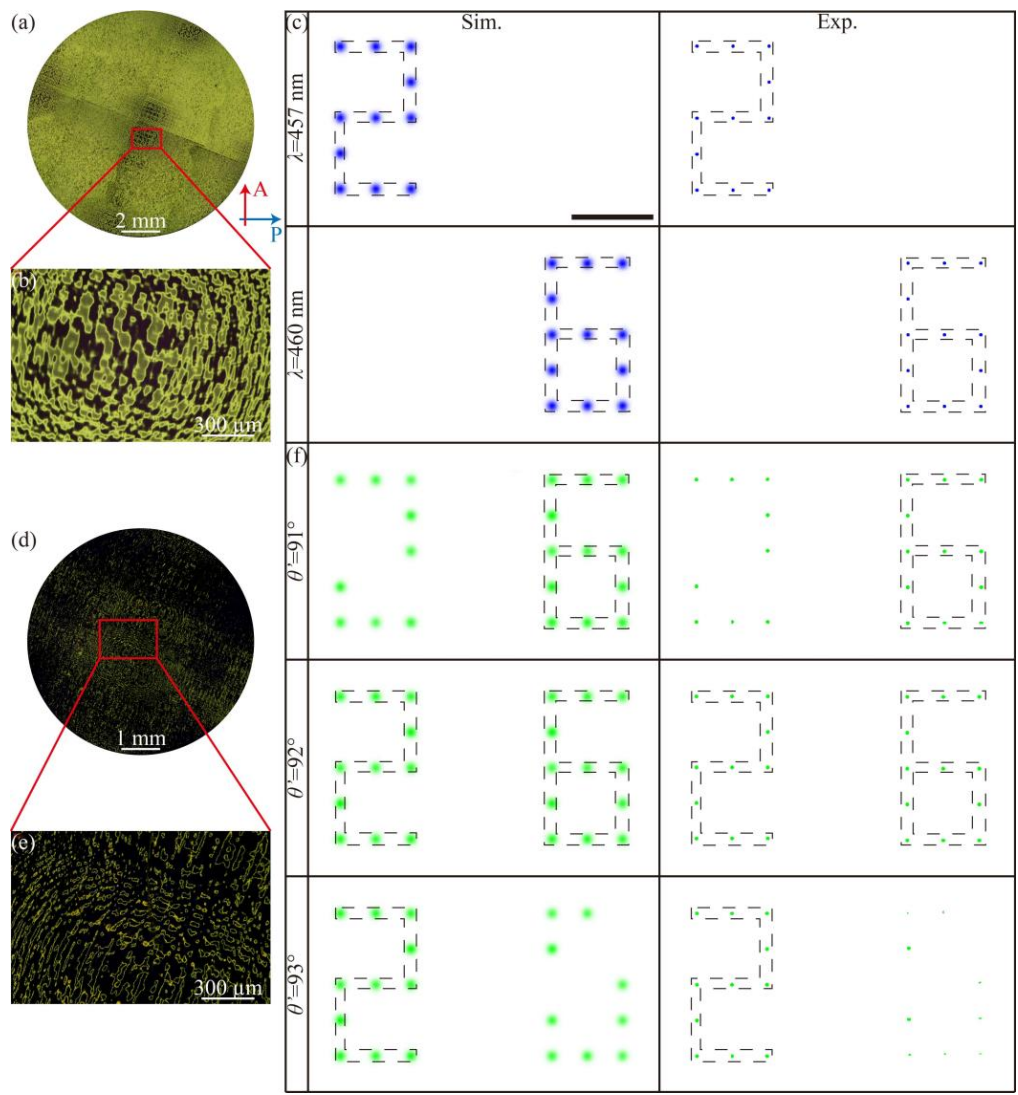

**Fig. 3** Holographic multiplexing accuracies with individual wavelengths and polarization states. (a)&(b) LC retarder for wavelengths only multiplexing, under polarization optical microscope, (c) simulation and experimental results for the incident wavelengths of 457 nm and 460 nm. (d)&(e) LC retarder for characterized polarization azimuthal angles only multiplexing, under polarization optical microscope, (f) simulation and experimental results for polarization-multiplexed hologram composed of 1° and 3° polarization angles, observed through analyzers set at 91°, 92°, and 93° polarization angles, respectively. The scale bar in Fig. 3(c) represents 0.5 mm.

The polarization states of the holographic patterns can be characterized using a retarder and a linear polarizer. Since these characterized polarization states serve as multiplexing parameters during the design of the LC retarder, this paper demonstrates the capability for fine multiplexing of holographic patterns at linear polarization azimuthal angles of 1°.

The LC ML, with a maximum radius of 2.916 mm and a focal length of 40 mm, is designed to produce two distinct holographic images corresponding to linear polarization states with angles of 1° and 3°. According to Malus's Law, the intensity of transmitted light varies as the square of the cosine of the angle between the planes of transmission. Consequently, the holographic patterns can be eliminated by an analyzer set at linear polarization azimuthal angles of 91° and 93°, respectively. However, when the polarization azimuthal angle is set to other angles, e.g.: 92°, both holographic patterns remain visible.

Fig. 3(d) shows the LC retarder designed for the finest differences in linear polarization azimuthal angles, as observed through a polarization optical microscope, with a microscale multi-domain LC alignment depicted in Fig. 3(e). Fig. 3(f) presents both simulation and experimental results for the characterized linear polarizer azimuthal angles of 91°, 92°, and 93°. Unlike the nearly completely separated holographic patterns observed with different incident wavelengths, after applying the same image processing procedures, it is noted that only two points of the holographic pattern "2" projected on the left side are eliminated by the linear analyzer set at 91°, while three points of the holographic pattern "6" projected on the right side are eliminated by the analyzer set at 93°. This crosstalk occurs due that other holographic points are not converted to perfect linearly polarized light, yet both holographic patterns can be clearly obtained with any other azimuthal angles. However, the experiment still demonstrates that the finest multiplexing of characterized polarization angle resolution is 2°, to reconstruct the entire holographic patterns.

Building on the aforementioned demonstrations of wavelength and polarization accuracy, we experimentally confirm in Supplementary Note 3 that the wavelength and polarization resolutions of multiplexed holograms of the same size, which carry six vortex wavefronts, are also 3 nm and 2°, respectively.

**Advanced holographic multiplexing with multiplexing wavefronts, wavelengths and polarization states**

To experimentally demonstrate our advanced holographic encoding methods with ultra-high capacity, we utilize six distinct topological charges: 5, 12, 20, 29, 39 and 50, corresponding to different multiplexing channels. The encoded amplitude values for $(A_{1,p1,q1,\lambda 1}, A_{2,p1,q1,\lambda 1}, \theta_{p1,q1,\lambda 1})$, $(A_{1,p2,q2,\lambda 1}, A_{2,p2,q2,\lambda 1}, \theta_{p2,q2,\lambda 1})$, $(A_{1,p3,q3,\lambda 1}, A_{2,p3,q3,\lambda 1}, \theta_{p3,q3,\lambda 1})$, $(A_{1,p4,q4,\lambda 1}, A_{2,p4,q4,\lambda 1}, \theta_{p4,q4,\lambda 1})$, $(A_{1,p5,q5,\lambda 1}, A_{2,p5,q5,\lambda 1}, \theta_{p5,q5,\lambda 1})$ and $(A_{1,p6,q6,\lambda 1}, A_{2,p6,q6,\lambda 1}, \theta_{p6,q6,\lambda 1})$ represent the following polarization states: (1, 1, 0) for linearly horizontal, (1, 1, π/2) for linearly vertical, (1, 0, 0) for left circularly polarized, (0, 1, 0), for right circular polarized (1, 1/2, 0) for left elliptically polarized and (1/2, 1, π/2) for right elliptically polarized states. This encoding method can also be applied for other incident wavelengths $\lambda_1$=470 nm, $\lambda_2$=520 nm, $\lambda_3$=550 nm, $\lambda_4$=590 nm, $\lambda_5$=650 nm and $\lambda_6$=730 nm, demonstrating the capability to manipulate the incident light across an additional degree of freedom. To achieve clear reconstructed images with this high capacity, the distance between adjacent foci *d* is set to 216 μm in this study.

The experimental optical diagram is shown in Supplementary Note 1. The supercontinuum laser (FL-SC-OEM from CNI, China), with wavelength range from 470 nm to 2400 nm, is expanded by a typical spatial filter and carries the desire wavelengths by inserting corresponding color filters. A spatial light modulator (FSLM-2K39-P02 from CAS Microstar, China) is employed to generate different OAM wavefronts carried by the laser beam. Fig. 4(a) displays the single LC retarder fabricated with the simulated phase profile to simultaneously regulate three dimensions: wavelength, wavefront, and polarization, as observed through a polarization optical microscope, with a microscale multi-domain LC alignment depicted in Fig. 4(b). The MLs on the LC retarder are designed with a maximum radius of 5.832 mm and a focal length of 90 mm, can generate six distinct polarization holographic images in the focal plane: *x*-linearly polarized, *y*-linearly polarized, left-circularly polarized, right-circularly polarized, left-elliptically polarized, and right-elliptically polarized, arranged from left to right when illuminated by incident vortex light of a single wavelength and topological charge. When illuminated by light with 6 different wavelengths (470 nm, 520 nm, 550 nm, 590 nm, 650 nm, and 730 nm) and 6 topological charges (-5, -12, -20, -29, -39, and -50), the LC retarder produces a total of 216 holographic patterns, which are further characterized by 6 different polarization states. The polarization states at various positions can be determined by measuring the Stokes vector (see Supplementary Note 2).

Due to the limitation of the manuscript, we only present holographic images captured under 2 topological charges (-5 and -20) and 6 different polarization states across the specified wavelengths as the demonstration in Figs. 4(c) and 4(d). We demonstrate 2-dimensional multiplexed holography based on polarization and wavelength using broadband polarization vortices, as detailed in Supplementary Note 4. Holographic images for other topological charges and wavelengths under varying polarization states are provided in Supplementary Note 5. The color holographic images with lower capacity, two polarizations and topological charge, and RGB triple-wavelength multiplexing are provided in Supplementary Note 5.

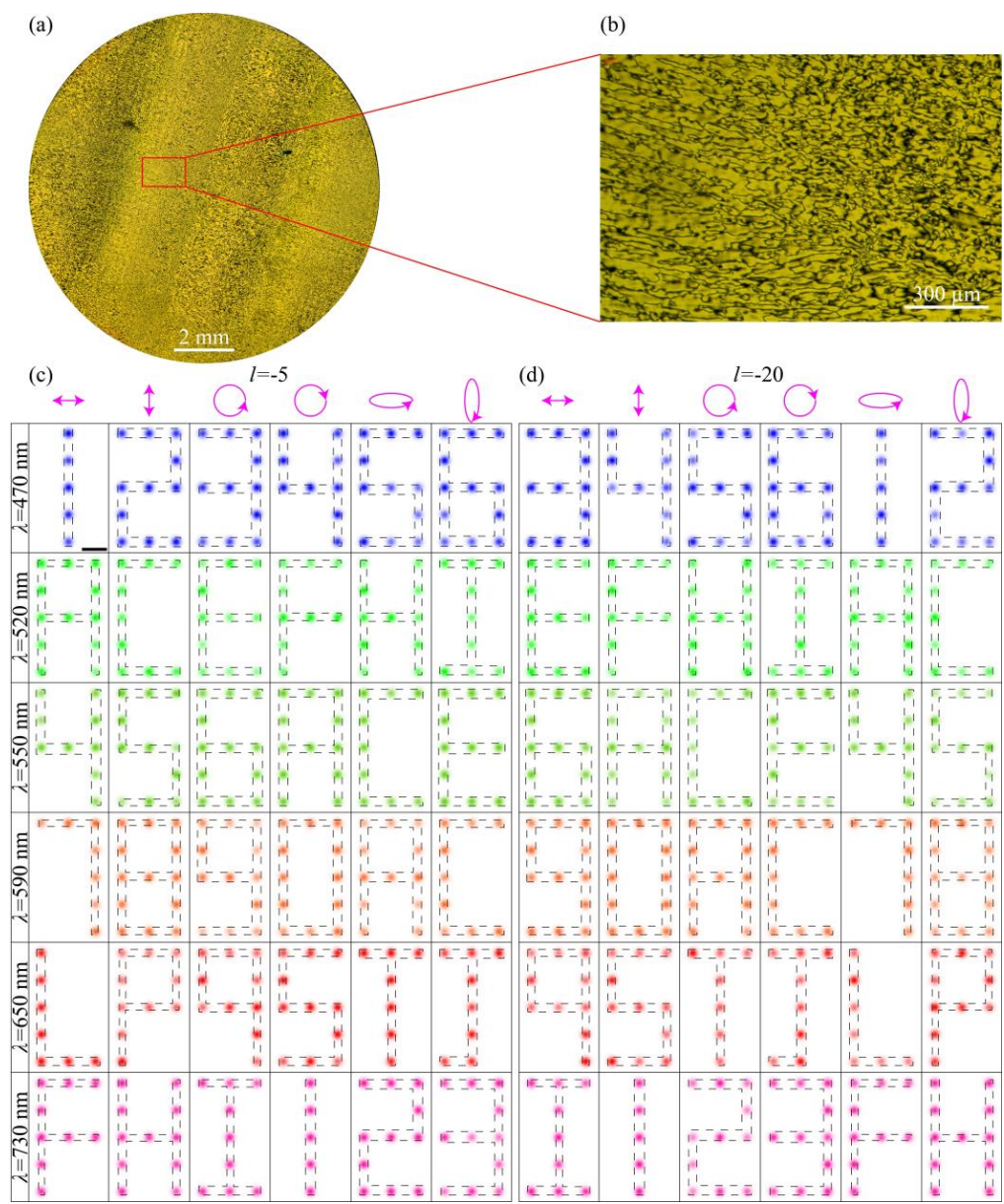

**Fig. 4** Advanced holographic multiplexing with multiplexing wavefronts, wavelengths and polarization states. (a)&(b) LC retarder under polarization optical microscope. Experimental

holographic patterns with incident vortex beam of (c) *l*=-5 and (d) *l*=20. The scale bar in Fig. 4(c) represents 200 μm.

When the incident vortex light has a topological charge of -5, Fig. 4(c) shows the experimentally measured holographic images ("1", "2", "3", "4", "5", "6"), ("A", "C", "E", "F", "H", "I"), ("4", "5", "6", "A", "C", "E"), ("7", "8", "9", "0", "A", "C"), ("L", "P", "Q", "S", "T", "J"), and ("F", "H", "I", "1", "2", "3") from top to bottom, corresponding to the incident wavelengths of 470 nm, 520 nm, 550 nm, 590 nm, 650 nm, and 730 nm, with the following polarization states: *x*-linearly polarized, *y*-linearly polarized, left-circularly polarized, right-circularly polarized, left-elliptically polarized, and right-elliptically polarized, respectively. Notably, during the actual experimental process, due to the extensive reuse of holographic channels by a single liquid crystal device (containing 216 holograms), and their larger area requiring assembly through stitching, this inevitably introduces certain stitching errors. The combined effect of these factors results in each hologram, after experimental measurement and filtering, having points representing letters composed of only a few pixels (in extreme cases, even a single pixel). This makes it extremely difficult to directly observe and identify the images with the naked eye when they are presented together in an abbreviated form. To address this issue, specific processing methods were applied to the measured holograms: first, a logarithmic function was used to adjust and ensure that the intensity differences between filtered points would not be excessively large; subsequently, a convolution operation using a Gaussian function was performed on each image, allowing each point to be displayed more clearly. This approach not only effectively demonstrated the functionality of the encoded device but also significantly improved analysis efficiency and accuracy without the need to individually magnify each holographic image for visual inspection. Similarly, when the incident vortex light has a topological charge of -20, Fig. 4(d) shows the experimentally measured holographic images ( "3", "4", "5", "6", "1", "2"), ("E", "F", "H", "I", "A", "C"), ("6", "A", "C", "E", "4", "5"), ("9", "0", "A", "C", "7", "8"), ("Q", "S", "T", "J", "L", "P"), and ("I", "1", "2", "3", "F", "H") from top to bottom, under the same incident wavelengths and polarization states. Thus, when vortex beams with varying wavelengths and topological charges illuminate the LC ML, holographic images with different polarization states are produced. This demonstrates the capability of a single LC retarder to simultaneously regulate multiple degrees of freedom, including wavelength, wavefront, and polarization.

**Discussion**

In summary, this study highlights the remarkable potential of conjugate phase encoding for multidimensional manipulation of optical field using liquid crystals. By integrating advanced polarization control with multiplexed holography across wavelengths, orbital angular momentums, and polarization states, we have successfully encoded 216 distinct holographic channels within a single photoinduced liquid crystal alignment unit. Our experimental results demonstrate impressive multiplexing accuracies of 3 nm in wavelength differences and 2 degrees in linear polarization azimuth differences, showcasing the precision and versatility of our approach.

The capabilities of conjugate phase encoding not only push the boundaries of ultra-high-capacity holographic information encoding but also pave the way for innovative applications in optical storage, communication, and encryption. Liquid crystal technology offers significant advantages over traditional optical components due to its ease of fabrication, high efficiency, broad bandwidth, cost-effectiveness, and multifunctionality. The straightforward manufacturing process and superior performance make it particularly well-suited for large-scale applications.

This research addresses the critical challenge of developing easily manufacturable, high-efficiency, low-cost, broadband, and high-capacity vector holography, thereby unlocking new possibilities for secure data encryption, high-bandwidth transmission, and sophisticated three-dimensional full-color displays. The liquid crystal orbital angular momentum multiplexed holography technique based on broadband multifocal spiral lens presented herein promises to significantly enhance these technologies, setting a new benchmark for intelligent and responsive photonic devices.

**Acknowledgments**

The research was financially supported by the Guangdong Major Project of Basic Research (2020B0301030009); National Natural Science Foundation of China (61935013, 62375181, 61975133); Science, Technology and Innovation Commission of Shenzhen Municipality (JCYJ20200109114018750); Shenzhen Peacock Plan (KQTD20170330110444030); Scientific Instrument Developing Project of Shenzhen University (2023YQ001); Shenzhen University 2035 Initiative (2023B004).

**Conflict of Interest**

The authors declare no conflict of interest.